# AI4AI: Quantitative Methods for Classifying Host Species from Avian Influenza DNA Sequence

Woo Yong Choi*, Kyu Ye Song, Chan Woo Lee
OrbisAI Inc., Seoul, South Korea

## Abstract

Avian Influenza breakouts cause millions of dollars in damage each year globally, especially in Asian countries such as China and South Korea. The impact magnitude of a breakout directly correlates to time required to fully understand the influenza virus, particularly the interspecies pathogenicity. The procedure requires laboratory tests that require resources typically lacking in a breakout emergency. In this study, we propose new quantitative methods utilizing machine learning and deep learning to correctly classify host species given raw DNA sequence data of the influenza virus, and provide probabilities for each classification. The best deep learning models achieve top-1 classification accuracy of 47%, and top-3 classification accuracy of 82%, on a dataset of 11 host species classes.

## 1. Introduction

Highly pathogenic avian influenza (HPAI) virus is extremely contagious among bird species and rapidly fatal, causing high mortality rates among domestic poultry and immense socioeconomic damage, especially in certain Asian countries such as China and South Korea. It often is initially spread in the form of lowly pathogenic avian influenza (LPAI) virus from wild birds and their feces to other bird species. Some strains of the LPAI virus infect certain bird species, and certain strains infect others, making its pathogenicity unpredictable and hard to track.

It is known that protein coding genomes of the Avian Influenza (AI) virus constitute a majority of the virus pathogenicity, yet predicting the effects of a genomic variation on the pathogenicity between other mammalian species remains a challenge. Current methodologies require animal testing of the newly obtained AI virus sample in different species. Such methodology requires more time and resources than is often allowed in a state of outbreak emergency. The virus sample needs to be sent to a laboratory, cultured, injected into animals, and observed for symptoms, requiring days and sometimes weeks.

To tackle this problem, we have developed a fully DNA sequence based, deep learning based approach to analyze DNA sequences *de novo*, and predict AI virus's pathogenicity across different mammalian species. Our deep learning model is based on convolutional neural networks (CNN), a powerful computer vision algorithm able to efficiently capture both obvious and latent features in an image. While computer vision tasks with convolutional neural networks deal with 2D images with 3 color channels, they can be adopted to DNA sequence data as well. The input sequence data will be rearranged into a 2D matrix with 4 channels representing the bases (A,T,C,G). When pixels with specific RGB values are oriented

---

Corresponding Author: cchoi@orbisai.co

in a certain way, they form a image feature such as a line, or a circle. When DNA sequences are ordered in a certain way, the relationship between sequences code a specific protein, which is a feature. Hence, determining the features and their impact on pathogenicity from this DNA 2D matrix is analogous to multi-class object detection tasks in computer vision.

CNNs have been successfully applied to predicting various effects of human DNA sequence variations in the past. For instance, DeepSEA is a deep learning model based on CNNs that predict the effects of noncoding DNA sequence variants of the human genome on TF-binding, DNA accessibility, and histone marks. DeepBind is also a CNN based model that predicts DNA sequence specificities for DNA and RNA binding.

Our model AI4AI, short for Artificial Intelligence for Avian Influenza, is inspired by DeepSEA and DeepBind, applying CNN based models to Avian Influenza virus DNA instead of the human genome to study how the coding regions impact interspecies pathogenicity.

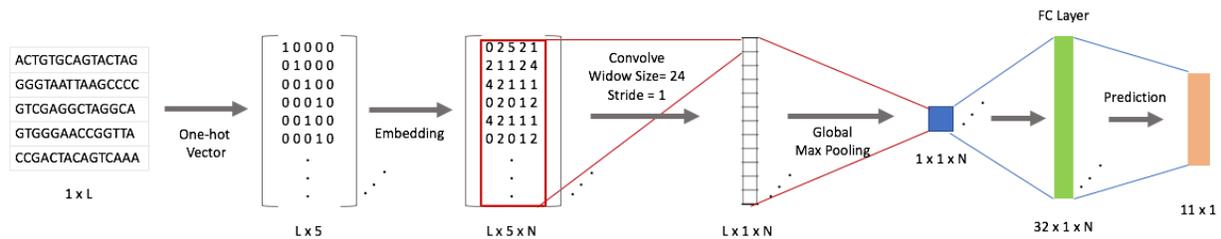

**Figure 1:** Data preprocessing begins with encoding DNA sequence data into a one hot vector, which is then randomly embedded into a L by 5 space. Basic architecture of convolutional neural networks comprises of first inputting data into a 1D convolutional layer with N filters, convolution window of 24 pixels and stride of 1 pixel. Then, the output vector of the convolution layer goes through global max pooling, fully connected layers, and finally prediction.

## 2. Methods

### 2.1 Data Acquisition and Preprocessing

Data used for training AI4AI prediction models were acquired from Influenza Research Database (https://www.fludb.org). DNA sequence data of Influenza A virus with H5 Clade classification, with all segments of the genome, and only complete genome submissions were collected as raw data, totaling 18,385 data entries and 121 host species classes.

This raw data is highly biased for only a small part of host species classes, hence is unfit for training deep learning models. Therefore, only the top most occurring 11 host species were considered in the final training set, totaling 14,810 entries.

The raw sequence data is then encoded to one-hot vector, where [A,C,T,G] is mapped to a one dimensional vector of length 5, last element being padding. For instance, A would be [1,0,0,0,0], and T would be [0,0,1,0,0], forming a vector dimension of Lx5, where L is the sequence length. For the purposes of this study, L was fixed at 2500 so that sequences longer than 2500 was clipped, and shorter than 2500 had padding appended to the end.

The sparsity of the resulting one-hot vector is high (a lot of zeroes), and may result in difficulties during training deep architectures. Thinking of it in terms of the human brain, this means that most neurons won't be connected to each other, causing the brain to hardly function. Therefore, we convert the one-hot vector into a randomly embedded vector.

We used tenfold cross validation across all models to obtain internal validation accuracies and errors. Data used for external validating and testing AI4AI models were acquired from Professor Song Chang Seon Lab in Konkuk University. It contained 82 entries with host species included in the 11 used in the training data.

## 2.2 Convolutional neural networks

At the most basic level, we developed a convolutional neural network as shown in Figure 1. After embedding, the data is duplicated N times, creating an input vector dimension of L x 5 x N. N is a hyperparameter indicating the number of learnable convolutional filters that will be applied to the input vector. For this study, N was chosen to be 128. In image object recognition, these learnable convolutional filters would form specific edges, curves, circles, or any other features that would make objects when used in combination with each other. In DNA sequence processing, each of these filters would in essence indicate a combination and permutation of DNA base sequences that may or may not create meaningful features, in this case interspecies pathogenicity, when merged in a certain way.

Each of these convolutional filters has a dimension of 24x1, and conducts 1-dimensional convolution with a stride of 1. This creates an output vector with a dimension of Lx1xN, which can be thought of as a *motif vector* containing important segments of the DNA sequence with respect to the interspecies pathogenicity.

In the next stage, the motif vector is inputted into a global max-pooling layer, where the strongest motif is chosen for inputting into the fully connected (FC) layer. The FC layer utilizes softmax classification layer at the end to assign each host species label a probability with respect to the input DNA sequence, as shown below where z is the input vector to the softmax layer, j index of output class labels, and K is the number of class labels:

$$\sigma(z)_j = \frac{e^{z_j}}{\sum_{k=1}^{K} e^{z_k}} \quad (1)$$

Based on this basic model, we have developed a variety of convolutional neural network based architectures as shown in Table 1.

| Model Name | Architecture |
|---|---|
| CNN | conv1d, maxpool, fc, softmax |
| CNN_BN | conv1, bn, maxpool, drop, fc, bn, drop, softmax |
| CNN+ | conv1, conv1d, conv1d, conv1d, maxpool, bn, conv1d, conv1d, maxpool, bn, conv1d, conv1d, conv1d, maxpool, bn, maxpool, drop, fc, drop, bn, drop, softmax |
| DenseNet | Standard DenseNet |
| ResNet | Standard Resnet |

**Table 1:** Deep learning architectures, bn: batch normalization, fc: fully connected, drop: dropout.

## 3. Results

We evaluate the five trained models as shown in Table 1 with tenfold cross validation on Influenza Research Database data.

Top-1 classification accuracy indicates the accuracy at which a model correctly predicts a host species with the highest output probability. Its results are shown in Table 2.

| Model Name | Accuracy |
|---|---|
| CNN | 47% |
| CNN_BN | 15% |
| CNN+ | 20% |
| DenseNet | 42% |
| ResNet | **59%** |

**Table 2:** Accuracies for each deep learning model

As seen in table 2, ResNet performed the best with Top-1 accuracy of 59%. CNN architectures with regularizations such as batch normalization and dropout performed extremely poorly, indicating underfitting of the model to the data. ResNet may have performed the best because it allows earlier convolutions of the data to have bigger influence in the prediction outcome than other models.

Using the best performing ResNet architecture, we conducted Top-3 accuracy tests using external data obtained from Konkuk University. Top-3 accuracy indicates the accuracy at which the label class is contained in the model prediction output with top 3 highest probabilities. The resulting accuracy levels reached as high as 82.43%. This metric is useful because AI virus strains not only infect one host species, but also other host species with a certain probability dependent on their DNA sequence. This Top-3 accuracy may be a benchmark for interspecies pathogenicity, and can be validated further as more external data is added to public databases.

To see how deep architectures compare against traditional machine learning algorithms, we ran random forest, gradient boosting, and xgboost algorithms on the same dataset for Top-1 classification accuracy. The results are shown in Table 3.

| Model Name | Accuracy |
|---|---|
| ResNet | 59% |
| Random Forest | **69%** |
| Gradient Boosting | 43% |
| xgboost | 54% |

**Table 3:** Traditional machine learning algorithms versus the best performing deep learning architecture.

Although ResNet performs better than both gradient boosting and xgboost, both based on decision trees, random forest performed with ~10% higher accuracy. Although deep learning architectures are known to perform well with high-dimensional feature space, they don't perform very well with small amount of data entries. On other hand, random forest and other machine learning models work better with smaller data, but also with smaller dimensional feature space. Perhaps in this case, the insufficient amount of data had a stronger influence on the performance of these models.

## 4. Future Work

This study has shown a new methodology of analyzing AI virus DNA sequence data *de novo*, using deep learning architectures. To improve the Top-1 accuracy of deep architectures, more open source training data may be required. In addition, Top-3 accuracy of these deep architectures will need to be further validated as we observe more interspecies infection of AI. Hence, to close, an elevated level of data submission participation from the community will be needed to improve this new quantitative methodology.

## Acknowledgements

We would like to thank Song Chang Seon lab for their support in providing us background information on Avian Influenza, as well as providing valuable validation dataset. We would also like to thank Ministry of Science, ICT and Future Planning of South Korea for funding this research.